# Detection of Coronavirus (COVID-19) Associated Pneumonia based on Generative Adversarial Networks and a Fine-Tuned Deep Transfer Learning Model using Chest X-ray Dataset

Nour Eldeen M. Khalifa[1,3 \[0000-0001-8614-9057\]] Mohamed Hamed N. Taha[1,3 \[0000-0003-0200-2918\]] Aboul Ella Hassanien[1,3 \[0000-0002-9989-6681\]] Sally Elghamrawy[2,3 \[0000-0002-5430-390X\]]

[1]Faculty of Computers and Artificial Intelligence, Cairo University, Giza, Egypt
[2]MISR Higher Institute for Engineering and Technology, Egypt, IEEE Member
[3]Scientific Research Group in Egypt (SRGE) http://www.egyptscience.net
[1]{nourmahmoud, mnasrtaha, aboitcairo}@cu.edu.eg
sally_elghamrawy@ieee.org

**Abstract.** The COVID-19 coronavirus is one of the devastating viruses according to the world health organization. This novel virus leads to pneumonia, which is an infection that inflames the lungs' air sacs of a human. One of the methods to detect those inflames is by using x-rays for the chest. In this paper, a pneumonia chest x-ray detection based on generative adversarial networks (GAN) with a fine-tuned deep transfer learning for a limited dataset will be presented. The use of GAN positively affects the proposed model robustness and made it immune to the overfitting problem and helps in generating more images from the dataset. The dataset used in this research consists of 5863 X-ray images with two categories: Normal and Pneumonia. This research uses only 10% of the dataset for training data and generates 90% of images using GAN to prove the efficiency of the proposed model. Through the paper, AlexNet, GoogLeNet, Squeeznet, and Resnet18 are selected as deep transfer learning models to detect the pneumonia from chest x-rays. Those models are selected based on their small number of layers on their architectures, which will reflect in reducing the complexity of the models and the consumed memory and time. Using a combination of GAN and deep transfer models proved it is efficiency according to testing accuracy measurement. The research concludes that the Resnet18 is the most appropriate deep transfer model according to testing accuracy measurement and achieved 99% with the other performance metrics such as precision, recall, and F1 score while using GAN as an image augmenter. Finally, a comparison result was carried out at the end of the research with related work which used the same dataset except that this research used only 10% of original dataset. The presented work achieved a superior result than the related work in terms of testing accuracy.

## 1. Introduction

The COVID-19 coronavirus [1] is one of the newest viruses on the earth which was announced in late December 2019. This new virus was declared as a pandemic by the World Health Organization which means the virus can geographically spread and affects an entire country or the whole world. This virus leads to pneumonia [2], which is an infection that inflames the lungs' air sacs [2]. One of the methods to detect those inflames is using X-rays for the chest. Here is the role of Artificial Intelligence and machine learning techniques would help doctors to detect pneumonia accurately and speedily.

Over the decade, machine learning technologies have been rapidly developed and integrated into CAD systems to provide accurate and rapid diagnosis. The remarkable success of Artificial Intelligence (AI) brings more signs of progress in medical image analysis. The ability of an effective AI model is highly dependent on learning from a sufficient amount of training samples [3].

Deep learning algorithms produce sophisticated results for different machine learning tasks and computer vision tasks. To perform well on a given task, these algorithms require a large data set for training. However, deep learning algorithms lack generalization and suffer from over-synthesis whenever they are trained in a small data set, especially when one deals with medical images.

For supervised image analysis in medical imaging, having image data along with their corresponding annotated ground-truths is costly and time-consuming since an-notations of the data is done by medical experts manually[4].

Convolutional neural networks (CNN) have shown advanced performance in various applications, thanks mainly to widely explained training data. Unfortunately, getting such huge medical annotations is a challenge. Classic data-augmentation techniques (DA) such as rotation are geometric/intensity modulations of original images for accurate diagnosis [5].

The remaining of this paper is organized as follows. In section 2, related work and scope work will be explored. In section 3, an overview of Generative Adversarial Networks and Deep Transfer Learning will be presented. In section 4, discusses the dataset used in the proposed model. In section 5, the proposed model's architecture will be presented while section 6 discusses our outcomes and discussion of the paper. Finally, section 7 provides conclusions and directions for further research.

## 2. Related Works

The coronavirus (Covid-19) [1] attracts the attention of many researchers to do further investigation about the symptoms of this viral disease. One of those ways of investigation is the detection of pneumonia from X-ray chest images. There a lot of datasets for chest X-rays for pneumonia such as [23]–[25], but in this research, the dataset in [25] has been selected due to the availability of data and the dataset has been used in many research to compare our work with as it will be presented in the next paragraphs.

D. S. Kermany et al in [26] presented medical diagnoses and treatable diseases by using image-based deep learning models to detect or classify different medical datasets including the dataset used in our research [25]. The testing accuracy for chest X-rays for pneumonia detection was 92.80 %. O. Stephen et al in [27] presented an efficient deep learning approach to pneumonia classification in healthcare based on the same dataset used in our research. The authors proposed a deep learning model with 4 convolutional layers and 2 dense layers in addition to classical image augmentation and achieved 93.73% testing accuracy.

Saraiva et al in [28] presented a classification of images of childhood pneumonia using convolutional neural networks. The authors proposed a deep learning model with 7 convolutional layers and 3 dense layers and achieved 95.30 % testing accuracy. G. Liang and L. Zheng in [29] presented a transfer learning method with a deep residual network for pediatric pneumonia diagnosis. The authors a deep learning model with 49 convolutional layers and 2 dense layers and achieved 96.70% testing accuracy. H. Wu et al in [30] presented a model to predict pneumonia with chest X-ray images based on convolutional deep neural learning networks and random forest, the authors achieved 97 % testing accuracy.

All the above-related works used the same dataset [25] that we used in this research. The main difference between our proposed model and other related works that according to literature surveys; this research considered one of the first trails to use the generative adversarial network to generate more images and make the proposed model immune from memorizing the dataset and overcome the overfitting problem.

## 3. Generative Adversarial Networks and Deep Transfer Learning

GANs are a special kind of neural network model where two networks are trained at once, one focusing on image generation and the other focusing on discrimination. Generative adversarial networks (GANs) provide a method for learning deep representations without widely explained training data [6]. These networks achieve learning by deriving reverse propagation signals through a competitive process involving a pair of networks [6]. The aggressive training scheme gained attention in both academia and industry because of its usefulness in facing field transformation, and its effectiveness in generating new image samples. GANs have made great progress and tremendous performance in many applications such as semantic image editing, pattern transfer, image synthesis, super-resolution and classification

**3.1 GAN architecture**

The main aspect of the GAN is a zero-sum game with a min-max of two people. In this game, one player benefits from the other player's equivalent loss. Here, players match different GAN networks called Mark and Generator. The main objective of discrimination is to determine whether the sample belongs to a false or true distribution [7]. Whereas, the generator aims to deceive the distinguished by generating a false sample distribution. A distinction produces opportunities or the probability that a particular sample will be a true sample. A high probability value shows that the sample is likely to be a true sample. A value near zero indicates that the sample is fake.

The probability value near 0.5 indicates an optimal solution so that the distinction cannot distinguish between true and false samples [7].

The general geometry of the GAN is shown in Figure 1. In a general architecture, the generative adversarial network contains two types of networks called discriminatorily and generator referred to as D and G respectively. GANs train a deep G-generated neural network that takes as inputs a random multidimensional sample z (from a Gaussian or standardized distribution) to generate a sample from the required distribution [7].

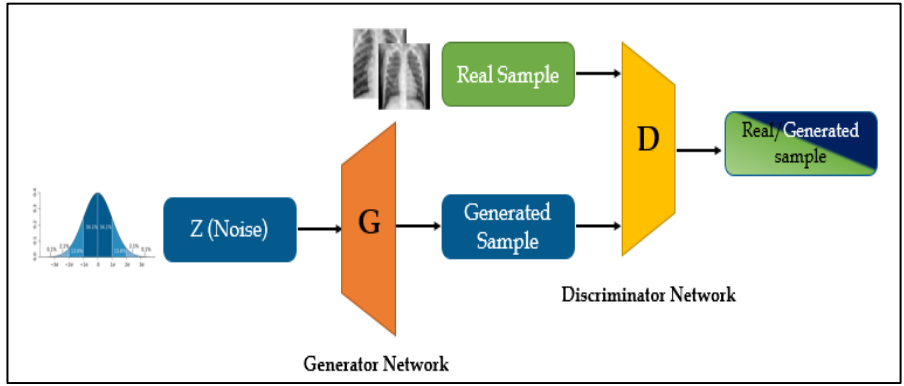

**Fig. 1.** Graphical representation of the generative adversarial network

**Generator**: The Generator (G), where G is a grid used to create images using random noise Z. Images generated using noise are recorded as G (z). Input, which is usually Gaussian noise, is a random point in the latent space. The parameters for both G and D networks are frequently updated during the GAN training process [8].

**Discriminator:** Discriminator is considered a discriminatory network to determine whether or not an image belongs to a true distribution. It receives an input X image and produces D (x) output, which represents the possibility that X belongs to a true distribution. If the result is 1, it indicates the true image distribution. The value of D output as 0 indicates that it belongs to a false image distribution [8]. The objective function of the min-imax game will be two players, as shown in Equation (1).

$$\text{Min}_G \text{Max}_D V(D,G) = E_{x \sim P_{data}(x)}[\log(d(x))] + E_{z \sim p_g(z)}[log(1 - D(G(z)))] \quad (1)$$

### 3.2 Deep Transfer Learning Networks

Deep learning, as a branch of artificial intelligence and machine learning, relies on algorithms to process data and simulate the thought process, or to develop abstractions [9] - [11]. Deep Learning (DL) assigns inputs to outputs using layers of methods to process and analyze hidden patterns in the data and visually expose things [12] - [14]. The data is passed through each layer of the deep network, with the output of the previous layer providing inputs for the next layer. The input layer is the first layer in the deep neural network, while the output layer is the final layer in the deep network. All hidden layers are between the input layers and the output layers [9], [15].

Years after, various advances in deep convolutional neural networks further reduced the error rate on the image classification competition tasks. CNN models demonstrated significant improvements in succeeding in the ImageNet Large Scale Visual Recognition Competition (ILSVRC) annual challenges. The Visual Geometry Group at Oxford (VGG) developed the VGG-16 and VGG-19 model for the ILSVRC-2014 competition with a 7.3% Top-5 error rate [16]. The winner of the ILSVRC 2014 competition was GoogleNet with a 6.7% Top-5 error rate [17]. In 2015, Residual Neural Network (ResNet) is the winner ILSVRC 2015 competition with a 3.6% Top-5 error rate. [18].A model called Xception [19] was introduced that uses depth-wise separable convolutions to outperform the Inception-V3 model [20] on the ImageNet [21] dataset classification task. Huang et. al introduced [22] A new CNN variant called Densely Connected Convolutional Networks (DenseNet) where each layer is connected directly to every later layer. The DenseNet has achieved considerable accuracy in classification, using significantly fewer computations and parameters, over the state-of-the-art.

## 4. Datasets characteristics

The pneumonia data set [25] used in this research was published in 2018 and is used for training, verification, and testing through this research. It contains 5,863 X-ray images (JPEG) and two categories: normal and pneumonia. A total of 5,863 patients had radiography from pediatric patients year to year from Guangzhou Medical Center. In this way, radiographs were performed as part of clinical care. All images in the dataset underwent treatment to remove all low-quality scans, as well as their classification by specialist doctors and by a third-party specialist, to prevent any classification error. This research used only 624 images selected from the dataset to demonstrate the efficacy of the proposed model that it could work with a limited number of images representing about 10.64% of the available data.

## 5. Proposed Model

The proposed model consists of three phases and based on two deep learning models. Figure 2 presents an abstract view of the proposed model. The first phase is the preprocessing phase which mainly is the augmentation process. The augmentation process, discussed in details in section 5.1, depends on the first deep learning model which is Generative Adversarial Networks (GAN). It is responsible for generating new images that will be used in the training and the testing phase. Figure 1 presents a graphical representation of the generative adversarial network used in this research. The proposed generative adversarial network consists of two main networks. The first network is generative, and the second network is the discriminator.

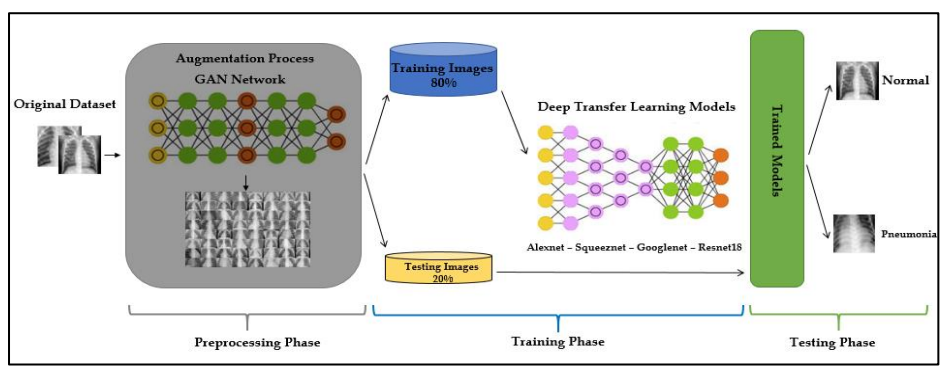

**Fig. 2.** Abstract view of the proposed model

The second phase is the training phase as illustrated in Figure 2. The training phase consists of first splitting the data into two parts and then training the deep transfer models. The first part is the training part which presents 80% of the dataset, while the second part is the testing part which presents 20% of the dataset

The authors of research tried first to build their deep neural networks based on the works presented in [31]–[33] but the testing accuracy wasn't acceptable. So, The proposed alternative way is to use deep transfer learning models to transfer the learning weights to reduce the training time, mathematical calculations and the consumption of the available hardware resources. This alternative method was adapted in similar research in [34], [35]. The selection of 80% for the training and 20% in testing proved it is efficient in many types of research such as [34], [35].

The second part of the training phase is the deep transfer models. The deep transfer models investigated in this research are AlexNet [36], SqueezeNet [37], GoogleNet [38], and ResNet18 [18]. The mentioned models have few layers when compared to large deep transfer models, such as Xception [19], DenseNet [22], and InceptionResNet [39], which consist of 71, 201 and 164 layers, respectively. The choice of these models reduces the training time and the complexity of the calculations. Figure 3 presents the number of layers of each deep transfer model along with the fine-tuned layer which is added in every model to adapt the number of classes in the dataset which contains 2 classes.

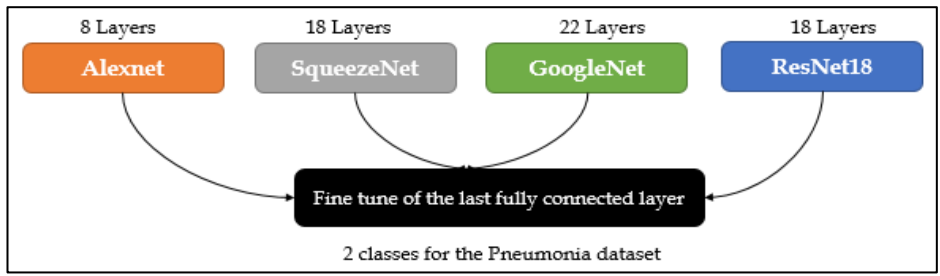

**Fig. 3.** The selected deep transfer models

The third phase of the proposed model is the testing phase as illustrated in Figure 2. The testing phase mainly deals with the testing dataset which contains 20% percent of the original dataset to evaluate every model depending on the testing accuracy and other performance metrics which will be presented in the following section.

The augmentation process of the proposed model is shown in Figure 4. It consists of three main phases: The first is the generator, the augmentation strategy and the third is the discriminator. The generator network consists of 5 transposed convolutional layers, 4 ReLU layers, 4 batch normalization layers, and Tanh Layer at the end of the model, while the discriminator network consists of 5 convolutional layers, 4 leaky ReLU, and 3 batch normalization layers. All the convolutional and transposed convoutional layers used the same window size of 4*4* pixel.

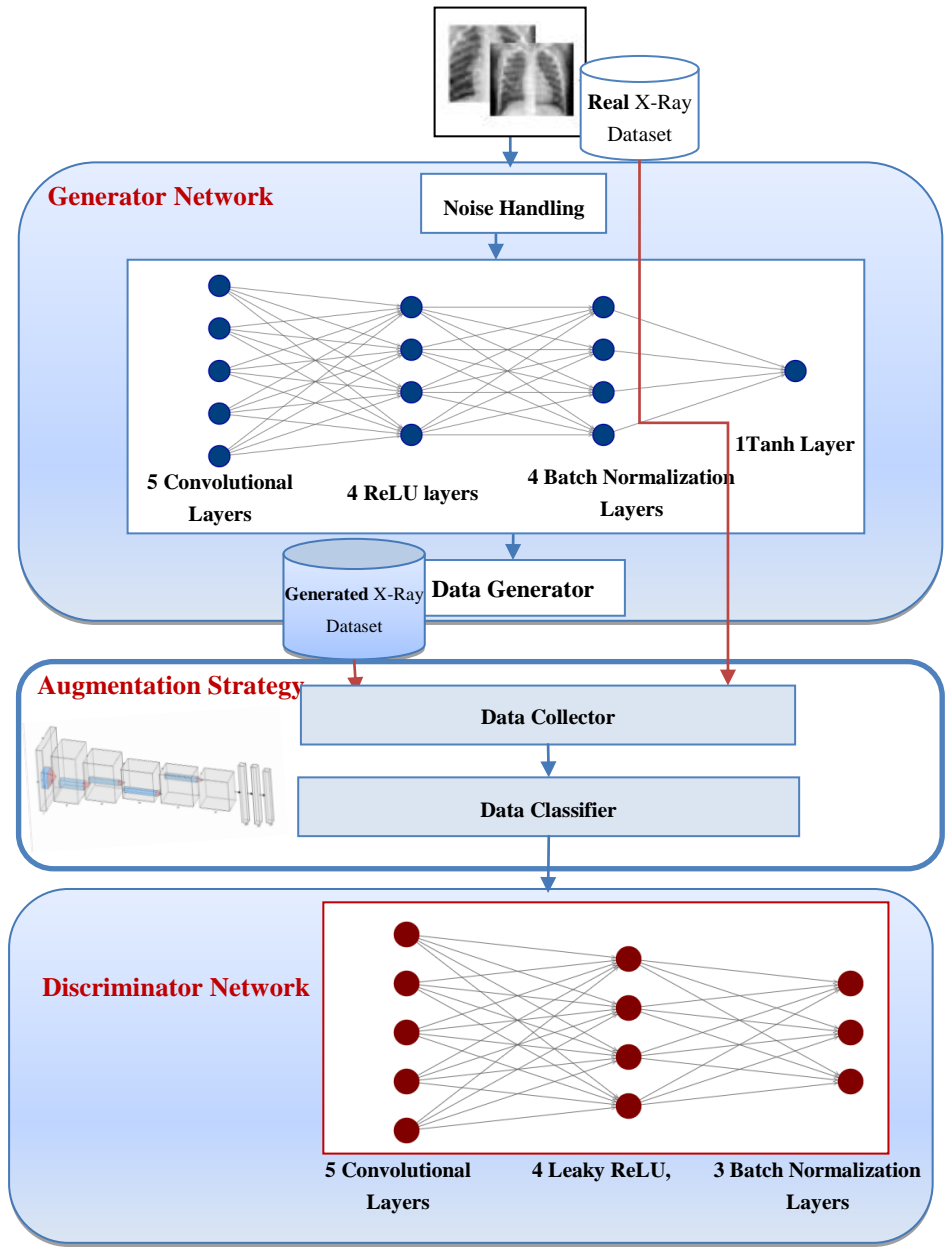

**Fig. 4.** The Augmentation Process in the Proposed Model

The structure of the generator and discriminator network is presented in Figure 5.

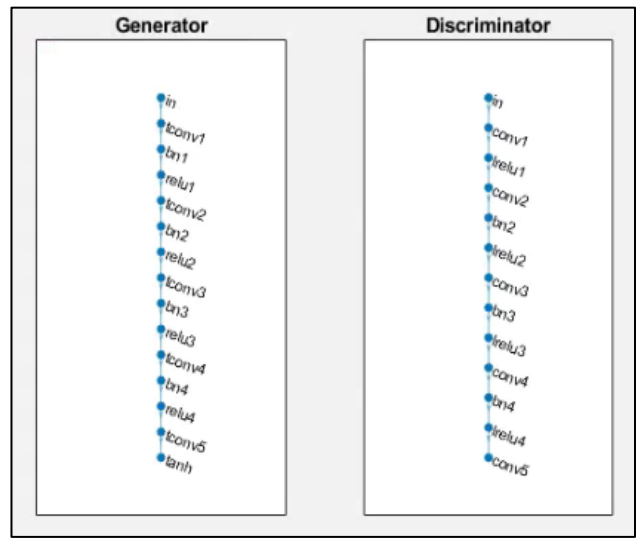

**Fig. 5.** The structure of generator and discriminator component

The GAN network helped in overcoming the overfitting problem caused by the limited number of images in the dataset. Moreover, it increased the dataset images to be 10 times larger than the original dataset. The dataset number of images reached 6,240 images after using the GAN network. This will help in achieving a remarkable testing accuracy and the performance matrices. The achieved results will be discussed in the experimental results section. Figure 6 presents a sample of the output of the GAN network.

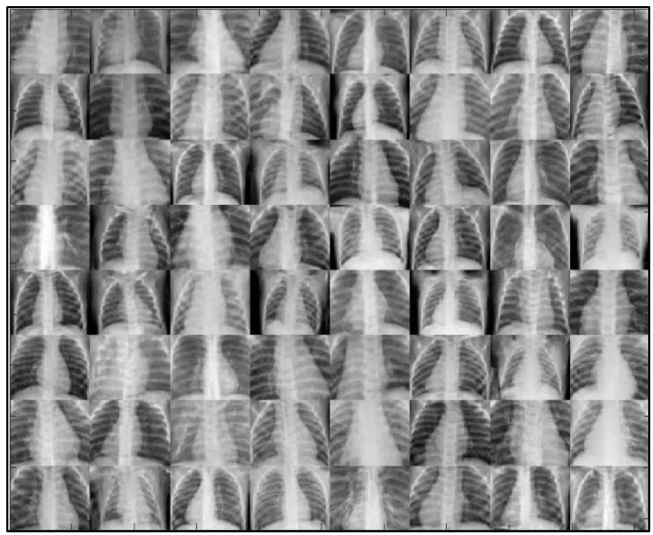

**Fig. 6.** Samples of generated images by GAN

## 6. Experimental Results

The proposed methodology was developed using a software package (MATLAB). The implementation was GPU specific. All experiments were performed on a computer with core i9 (2 GHz), 32 GB of RAM with Titan X GPU.

All experiment trials were carried out with only 10% percent of the original dataset while generating the other 90% using GAN. Three experimental trials were carried out with different 10% of the original dataset. The average of these trails was calculated and presented in the following subsections.

### 6.1 Confusion Matrix and Testing Accuracy

Testing accuracy is an estimation that demonstrates the precision and accuracy of any of the selected deep transfer models. Additionally, the confusion matrix is an accurate measurement that provides more insight regarding the achieved testing accuracy for every class. Figures 5,6,7 and 8 present the confusion matrices for the distinctive deep transfer models used in this research without/with using GAN. The Figures (7,8,9 and 10) illustrate that using GAN to generate new images and the train the deep transfer model achieves better testing accuracy. The testing accuracy was calculated using equation (2). The achieved testing accuracy for AlexNet, Squeezed, GoogLeNet, and Resnet18 using GAN is 96.1%, 94.7%, 98.6%, and 99% consecutively. The Resnet19 achieved the highest testing accuracy with 99% followed by GoogLeNet, AlexNet, and Squeezed.

One of the advantages of generating a confusion matrix is the measurement of testing accuracy for every class. Table 1 present the testing accuracy for every class using GAN for the different deep transfer models. Table 1 illustrates that the highest testing accuracy for the class "Normal" achieved by squeezenet model with 99.3%, while the highest testing accuracy for the class "Pneumonia" achieved by GoogLeNet model with 99.4%. Furthermore, the highest overall testing accuracy to distinct between the two-class achieved by Resnet18 with 99%.

**Fig. 7.** Confusion matrix for Alexnet (a) without GAN, and (b) with GAN

**Fig. 8.** Confusion matrix for Squeezed (a) without GAN, and (b) with GAN

**Fig. 9.** Confusion matrix for Googlenet (a) without GAN, and (b) with GAN

**Fig. 10.** Confusion matrix for Resnet (a) without GAN, and (b) with GAN

Thus, the Resnet18 was selected in our proposed model as the overall testing accuracy is the chosen metric to judge between the different deep transfer model besides the performance metrics discussed in the following subsection.

Table 1. Testing accuracy for every class for the different deep transfer models using GAN

| Accuracy/Model | AlexNet | Squeeznet | GoogLeNet | Resnet18 |
|---|---|---|---|---|
| **Normal** | 91.8% | 99.3% | 97.3% | 98.7% |
| **Pneumonia** | 98.9% | 92.5% | 99.4% | 99.2% |
| **Total Accuracy** | 96.1% | 97.8% | 96.8% | **99.0%** |

### 6.2 Performance Evaluation

To evaluate the performance of the deep transfer models, more performance matrices are needed to be investigated through this research. The most common performance measures in the field of deep learning are Precision, Recall, and F1 Score [40], and they are presented from equation (2) to equation (5).

$$\text{Testing Accuracy} = \frac{(TN+TP)}{(TN+TP+FN+FP)} \quad (2)$$

$$\text{Precision} = \frac{TP}{(TP+FP)} \quad (3)$$

$$\text{Recall} = \frac{TP}{(TP+FN)} \quad (4)$$

$$\text{F1 Score} = 2 * \frac{\text{Precision} * \text{Recall}}{(\text{Precision}+\text{Recall})} \quad (5)$$

Where TP is the count of True Positive samples, TN is the count of True Negative samples, FP is the count of False Positive samples, and FN is the count of False Negative samples from a confusion matrix.

Table 2 presents the performance metrics for the different deep transfer models. The table illustrates that the Resnet model achieved the highest percentage for precision, recall, and F1 score metrics with a percentage of 98.97%. The achieved percentage in all performance metrics concludes that Resnet18 is the optimal choice to be used in our proposed model along with using GAN to generate more data which helped in achieving the previous results and strengthen by the performance matrices.

Table 2. Performance matrices for the different deep transfer models using GAN

| Metric/Model | Alexnet | Squeeznet | Googlenet | Resnet18 |
|---|---|---|---|---|
| **Precision** | 96.52% | 93.60% | 98.63% | 98.97% |
| **Recall** | 95.37% | 95.88% | 98.31% | 98.97% |
| **F1 Score** | 95.94% | 94.46% | 98.47% | 98.97% |

## 6.3 Comparative Analysis

The dataset used in this research [25] was published in 2018 and it has been used in much research since that time. Table 3 illustrates the related works with the achieved testing accuracy. It is shown clearly that the proposed model using GAN and Resnet18 achieved a superior testing accuracy for all the related works.

Table 3. The comparative result with related works.

|  | Year | Description | Testing Accuracy |
|---|---|---|---|
| [26] | 2018 | Convolutional Neural Network (CNN) | 92.80% |
| [27] | 2019 | Deep learning model with 4 convolutional layers and 2 dense layers + classical Augmentation | 93.73% |
| [28] | 2019 | Deep learning model with 7 convolutional layers and 3 dense layers | 95.30 % |
| [29] | 2019 | Deep learning model with 49 convolutional layers and 2 dense layers | 96.70% |
| [30] | 2020 | Convolutional Neural Network (CNN) + Random forest | 97.00% |
| Proposed Model | 2020 | GAN + Resnet18 | 99.00% |

## 7 Conclusions and Future Works

The COVID-19 coronavirus is one of the newest viruses on earth which was announced in late December 2019. This virus leads to pneumonia, which is an infection that inflames your lungs' air sacs. One of the methods to detect those inflames is by using X-rays for the chest. In this paper, a pneumonia chest x-ray detection based on generative adversarial networks (GAN), and a fine-tuned deep transfer learning for a limited dataset will be presented. The use of GAN made the proposed model robust and overcome the overfitting problem and generated more images from the dataset. The dataset set used in this research consisted of 5863 X-ray images with two categories: Normal and Pneumonia. This research used only 10% of the dataset for training data and generated 90% of images using GAN to prove the efficiency of the proposed model. Through the paper, Alexnet, Googlenet, Squeeznet, and Resnet18 were selected as deep transfer learning models. Those models were selected based on their small number of layers on their architectures, which will reflect in reducing the complexity of the models and the consumed memory and time. Using a combination of GAN and deep transfer models proved it is efficiency according to testing accuracy metric. The research concludes that Resnet18 is the most appropriate deep transfer model according to testing accuracy measurement and achieved 99% with using GAN as an image augmenter. Finally, a comparison results were carried out at the end of the research with related work which used the same dataset except that this research used only 19% of it. The presented work achieved a superior result than the related work in terms of testing accuracy.

## Acknowledgments

We gratefully acknowledge the support of NVIDIA Corporation, which donated the Titan X GPU used in this research.